Research article

# Relationship between metabolic and genomic diversity in sesame (*Sesamum indicum* L.)

Hernán Laurentin[1], Astrid Ratzinger[2] and Petr Karlovsky*[2]

Address: [1]Universidad Centroccidental Lisandro Alvarado, Biologic Sciences Department, Agronomy Faculty, Barquisimeto, Venezuela and [2]University Goettingen, Molecular Phytopathology and Mycotoxin Research Unit, Goettingen, Germany

Email: Hernán Laurentin - hlaurentin@ucla.edu.ve; Astrid Ratzinger - aratzin@gwdg.de; Petr Karlovsky* - pkarlov@gwdg.de

* Corresponding author

## Abstract

**Background:** Diversity estimates in cultivated plants provide a rationale for conservation strategies and support the selection of starting material for breeding programs. Diversity measures applied to crops usually have been limited to the assessment of genome polymorphism at the DNA level. Occasionally, selected morphological features are recorded and the content of key chemical constituents determined, but unbiased and comprehensive chemical phenotypes have not been included systematically in diversity surveys. Our objective in this study was to assess metabolic diversity in sesame by nontargeted metabolic profiling and elucidate the relationship between metabolic and genome diversity in this crop.

**Results:** Ten sesame accessions were selected that represent most of the genome diversity of sesame grown in India, Western Asia, Sudan and Venezuela based on previous AFLP studies. Ethanolic seed extracts were separated by HPLC, metabolites were ionized by positive and negative electrospray and ions were detected with an ion trap mass spectrometer in full-scan mode for m/z from 50 to 1000. Genome diversity was determined by Amplified Fragment Length Polymorphism (AFLP) using eight primer pair combinations. The relationship between biodiversity at the genome and at the metabolome levels was assessed by correlation analysis and multivariate statistics.

**Conclusion:** Patterns of diversity at the genomic and metabolic levels differed, indicating that selection played a significant role in the evolution of metabolic diversity in sesame. This result implies that when used for the selection of genotypes in breeding and conservation, diversity assessment based on neutral DNA markers should be complemented with metabolic profiles. We hypothesize that this applies to all crops with a long history of domestication that possess commercially relevant traits affected by chemical phenotypes.

## Background

The diversity of characters among members of a species is an inherent feature of biological complexity. Most studies of biological diversity in crops have focused on morphological characters and DNA markers, covering both ends of the path of gene expression from genome to phenotype. Genome analysis records and compares the genetic make-up of lineages or individuals based on DNA sequences or fragment patterns. Both sequence analysis and DNA fingerprinting sample genome diversity, which is independent of environmental conditions and the developmental stage of the organism [1]. AFLP markers [2] are anonymous and are generally thought to be selectively neutral, which probably holds true for many kinds of DNA markers [3]. Even whole-genome sequencing of populations, the ultimate genome diversity survey tool, reveals at most the potential of a population to express various phenotypic features. Approaches based on transcriptomics and proteomics can identify gene expression patterns that underlie the current phenotype and that are affected by environment and the developmental stage of the organism. The relationship between the abundance of mRNA and protein molecules on one side and of phenotypic features relevant for crop production on the other is obscure and cannot yet be exploited for breeding purposes even in major crops with extensive genomic resources, let alone in minor or orphan crops.

A third level of gene expression, represented by the metabolic constitution of the organism, is directly related to features that are important in plant production. We are interested in secondary metabolites, because these natural products provide most of the chemical diversity in plants, and are a key factor (i) affecting the resistance of crops to pathogens and pests, and (ii) controlling commercially relevant traits such as taste, color, aroma and antioxidative properties.

The metabolic phenotype of an organism is analyzed by metabolomics, whose final goal is to identify and quantify all of the metabolites present in a sample [4,5]. Such a complete inventory is not attainable with current technology even for model organisms, so different types of metabolite analysis with more limited scopes serve as surrogates. Metabolic fingerprints are a static set of analytical signals originating from small molecules (e.g. HPLC peaks, TLC spots, or mass spectra), which can be used for diagnostic purposes or to confirm the origin of a sample. In metabolic profiling, which is analogous to transcription profiling, metabolic signals, either anonymous or assigned to structures, are generated and evaluated quantitatively for samples originating from different varieties, physiological states or treatments. Term profiling is also used for a comprehensive analysis of a class of substances defined by common structural features (e.g., oxylipin profiling). Alternative definitions of metabolic profiling and fingerprinting [6,7] are likely to lead to confusions whenever metabolic analysis and genome fingerprinting are treated jointly.

Sesame (*Sesamum indicum* L.) is one of the most ancient crops [8,9]. Sesame seed is highly nutritive (50% oil and 25% protein) and may be consumed directly or pressed to five an oil of excellent quality. Most studies of secondary metabolites in sesame focused on the lignans sesamin, sesamol, sesamolin and sesaminol [10-13] in seeds. These natural products have antioxidative properties and may confer health-promoting qualities on products containing sesame seeds or oil [14-17]. Sesame lignans also may play a role in the resistance of sesame to insect pests and microbial pathogens [18-23]. The metabolism of sesame lignans after ingestion is understood to a limited extent [24]. Metabolic profiling has not been a part of diversity studies in sesame.

Our objective in this study was to compare metabolic and genomic diversity in sesame and to discern the relationship between the two sets of data. Based on the difference in the diversification of sesame at the genomic and metabolic levels we will assess the usefulness of metabolic profiles in the identification of parent lines for breeding programs and in the selection of accessions for biodiversity preservation in sesame.

**Results**
Sesame accessions for this study were selected based on previously published AFLP data and represent most of the genome diversity in sesame from India, Western Asia, Sudan and Venezuela. Among these accessions, eight accessions have Jaccard similarity coefficients from pairwise comparisons that range from 0.39 to 0.85. These accessions encompass nearly all of the genome diversity detected by AFLP in the two-dimensional space of principal coordinate analysis and represent the four previously described major clusters [25]. The two Venezuelan genotypes, an experimental line and a commercial cultivar, were included because they represent Venezuelan breeding products with a Jaccard's similarity coefficient of 0.45 [26]. These genotypes represent the two major clusters comprising Venezuelan commercial cultivars and contain 80% of the total genetic diversity of sesame in Venezuela.

Three hundred and eighty one AFLP markers, ranging from 100 to 550 base-pairs, were recorded using 8 primer combinations. Ninety-five percent of the markers were polymorphic. Eighty-eight bands (23%) were unique, ranging from 5 (Turkey) to 21 (India 7) per accession (Table 1).

The reproducibility of the metabolic analysis was very good because similarity and dissimilarity measures and principal component analysis results showed negligible differences regarding three independent profiles generated from extracts of Sudan3 accession and compared to the other 9 accessions. The average of the three replicas obtained for Sudan3 was used for all further analysis.

Eighty-eight dominant metabolic signals were selected based on the mass chromatogram quality index, 47 of them in negative mode ESI and 41 in positive mode ESI.

BMC Genomics 2008, **9**:250                                                                                     http://www.biomedcentral.com/1471-2164/9/250**Table 1: AFLP: Primer combinations and polymorphism of DNA bands**

| Primer combination | Bands total | Polymorphic bands | Unique bands | | | | | | | |
|---|---|---|---|---|---|---|---|---|---|---|
| (Cy5)E_ACA+M_CAA | 55 | 51 | India1 1 | India5 1 | Turkey 4 | Sudan3 2 | Inamar 1 | | 43×32 1 | Total 10 |
| (Cy5)E_ACA+M_CAC | 53 | 50 | India1 1 | India7 2 | India8 4 | Sudan2 2 | Inamar 3 | | 43×32 1 | Total 13 |
| (Cy5)E_ACA+M_CAG | 51 | 49 | | India1 3 | | | Syria 1 | | | Total 4 |
| (Cy5)E_ACA+M_CAT | 78 | 76 | India1 2 | India5 1 | India7 1 | Syria 5 | Inamar 1 | | 43×32 1 | Total 11 |
| (Cy5)E_ACA+M_CCA | 41 | 38 | India7 8 | | Inamar 1 | | | | 43×32 2 | Total 11 |
| (Cy5)E_ACA+M_CCC | 37 | 36 | India1 1 | India5 4 | India7 2 | India8 1 | Turkey 1 | Sudan2 1 | Inamar 5 | 43×32 1 | Total 16 |
| (Cy5)E_ACA+M_CGAA | 25 | 25 | India5 1 | | India7 6 | | Syria 2 | Sudan2 3 | Inamar 1 | 43×32 2 | Total 15 |
| (Cy5)E_ACA+M_CTCA | 41 | 38 | India7 2 | | India8 4 | | Syria 1 | | | 43×32 1 | Total 8 |

**Table 2: Metabolic signals in sesame seed extracts used in the analysis**

| | Negative mode | | |
|---|---|---|---|
| Mass range | m/z values total | m/z values common to all accessions | Number of accession-specific m/z values or values lacking in only one accession |
| 200–400 | 14 | 3 | 2 (Sudan 2) — Total: 2 |
| 400–600 | 11 | 1 | 1 (Turkey) — Total: 1 |
| 600–800 | 17 | 10 | 1 (India 5) — Total: 1 |
| 800–1000 | 5 | 4 | 1 (India 5) — Total: 1 |
| | Positive mode | | |
| Mass range | m/z values total | m/z values common to all accessions | Number of accession-specific m/z values or values lacking in only one accession |
| 400–600 | 2 | 1 | 0 |
| 600–800 | 25 | 6 | 1 (India 8)   1 (43 × 32) — Total: 2 |
| 800–1000 | 14 | 9 | 1 (India 5) — Total: 1 |

More than 50% of the signals resulted from peaks eluting in a well-resolved area with retention times between 15 and 27 min. Thirty-four signals were common to all accessions, 16 in positive mode ESI and 18 in negative mode ESI. Eight signals were either accession-specific or present in all except one accession (Table 2). No association was found between the distribution of unique AFLP markers and accession-specific metabolic signals.

The coefficient of correlation between correlation coefficient-based similarity matrix and simple-matching coefficient-based similarity matrix was 0.63 (P < 0.01). Correlation between matrices obtained from AFLP data and metabolic profiles was not significant. Comparisons of matrices of metabolic data with Jaccard's coefficient matrix of AFLP data resulted in a correlation coefficient of -0.09 (P < 0.33) for the simple matching coefficient matrix and -0.24 (P < 0.18) for the correlation matrix. There were consistencies in scatter plots for pairs of accessions that fell into the same category (high similarity, intermediate similarity, or low similarity) for genomic and metabolic data (Fig. 1). Accession pair Syria-Sudan3 had high similarity on both axes, Sudan2–43 × 32, India7–43 × 32, and India5-Sudan2 were dissimilar both in metabolic profiles and AFLP fingerprints and pairs India1-India8, India1-Syria, India1-Turkey, India1-43 × 32 and India8-Turkey had intermediate similarities.

Biplot of principal coordinate analysis based on AFLP data calculated from Jaccard's coefficient captured 64% of the total variation (Fig. 2). Accessions Sudan2 and India7 on one side, and commercial cultivars Inamar and 43 × 32

Page 3 of 11
*(page number not for citation purposes)*

on the other, are the most distinctive. Biplots of principal component/coordinate analysis based on correlation coefficient, which captured 62% of the variation, and simple matching coefficient, which captured 77% of the variation, had similar patterns in that accessions India5 and 43 × 32 formed one group and the remaining eight accessions formed a second group. Visual comparison of biplots obtained with AFLP and metabolic profiles confirmed the classification of cultivar 43 × 32 as the most distinctive, which explains why Sudan2-43 × 32 was one of the most dissimilar pairs. Both biplots grouped Syria-Sudan3 as the most similar pair and India1-Syria, India1-Sudan3, India1-Turkey and I8-43 × 32 as pairs with intermediate similarities. The most important contradiction between both biplots was the placement of India5, India7 and Inamar. Based on AFLPs, India7 and Inamar were the most distinctive accessions, whereas metabolic profiles grouped them together with six other accessions. The opposite situation was found for India5, which was classified as one of the most distinctive based on metabolic profiles, but groups together with 5 other accessions based on the AFLP data.

## Discussion

Seed metabolic profiles were unrelated to the geographic origin of the accessions studied, which is similar to results obtained previously for genome diversity as assessed by AFLPs [25]. The relationship patterns generated for the AFLP data and for the seed metabolic profiles were different. No relevant data from other plant species are available for comparison, but there are two studies of the relationship between genomic and metabolic diversity in microorganisms. In rhizobia (bacteria), metabolic and genomic data (AFLP) were unrelated [27], while there were strong similarities between genome variation and metabolite diversity between two endophytic fungi [28].

If the number of characters reflects the sampling depth, then the metabolic profiles and AFLP fingerprints cover only a small portion of the underlying character sets. The AFLP-based analysis appears more robust because it was based on 363 polymorphic bands while only 88 signals were evaluated in the metabolic profiles. However, the metabolic profiles may contain more information because they are based on continuous rather than binary variables. To test this hypothesis we transformed the metabolic data into a binary matrix and compared the binary and continuous results. The quantitative information (normalized amplitudes of mass signals) did not affect the similarity patterns and therefore can be neglected in diversity surveys.

Diversity in AFLP patterns and metabolic profiles reflect different facets of genomic polymorphism. AFLPs are insensitive to gene expression and may occur most frequently in noncoding portions of the genome. Seed metabolic profiles result from biosynthetic activities in embryo and endosperm based on the expression of a small fraction of the total genomes. If the samples are representative, then differences between the diversity patterns are due to differences in the diversification of sesame at genomic and metabolic levels. Because the majority of plant genomes consist of noncoding sequences, most changes in AFLP patterns are expected to result from neutral mutations fixed by genetic drift rather than by selection. On the other hand all metabolites synthesized by a plant affect its fitness: apart from the metabolic costs incurred, anabolic processes are subjected to different selection pressures, both positive (e.g., resistance to pathogens, protection against light, improved dissemination of seeds) and negative (e.g., reduced attractiveness of seeds for disseminating animals because of a bitter taste, volatiles attracting pests, trigger of the germination of microbial pathogens). The synthesis of many secondary metabolites is known to be limited to conditions under which they enhance the fitness of their producer, limiting the costs of biosynthesis [41]. Metabolic profiles of sesame recorded under different environmental conditions are therefore likely to differ. For example, exposure to biotic stress is likely to generate defence-related signals, which may not be present in metabolic profiles of plants grown in the absence of pathogens and pests. Regardless of the progress in analytical technologies, chemical diversity revealed by metabolic profiling under a single set of conditions therefore remains an underestimate of the total metabolic capacity of sesame.

The genetic basis of the variation in the metabolic composition on plants was proven by the association between metabolic peaks detected by HPLC-MS and specific genomic loci in segregating populations of *A. thaliana* [29]. The disparity between the diversity patterns represented by AFLPs and by metabolic profiles thus provides insights into the processes that led to the composition of the current sesame genome. With the growing availability of instrumentation and software tools for nontargeted metabolite analysis by HPLC-MS [30,31], large-scale metabolic profiling is becoming a feasible task for diversity studies in cultivated plants. From a practical point of view, crop improvement programs [32] will benefit from the complementation of diversity assessment based on DNA markers by metabolic profiling particularly for plants such as pepper [35,40], mulberry [36,37,39] and fenugreek [38], the commercial value of which is largely affected by complex mixtures of secondary metabolites.

In addition to genuine differences in similarity patterns between genomic and metabolomic profiles caused by differences in diversification rates, non-representative sampling also may lead to inconsistencies. The involve-

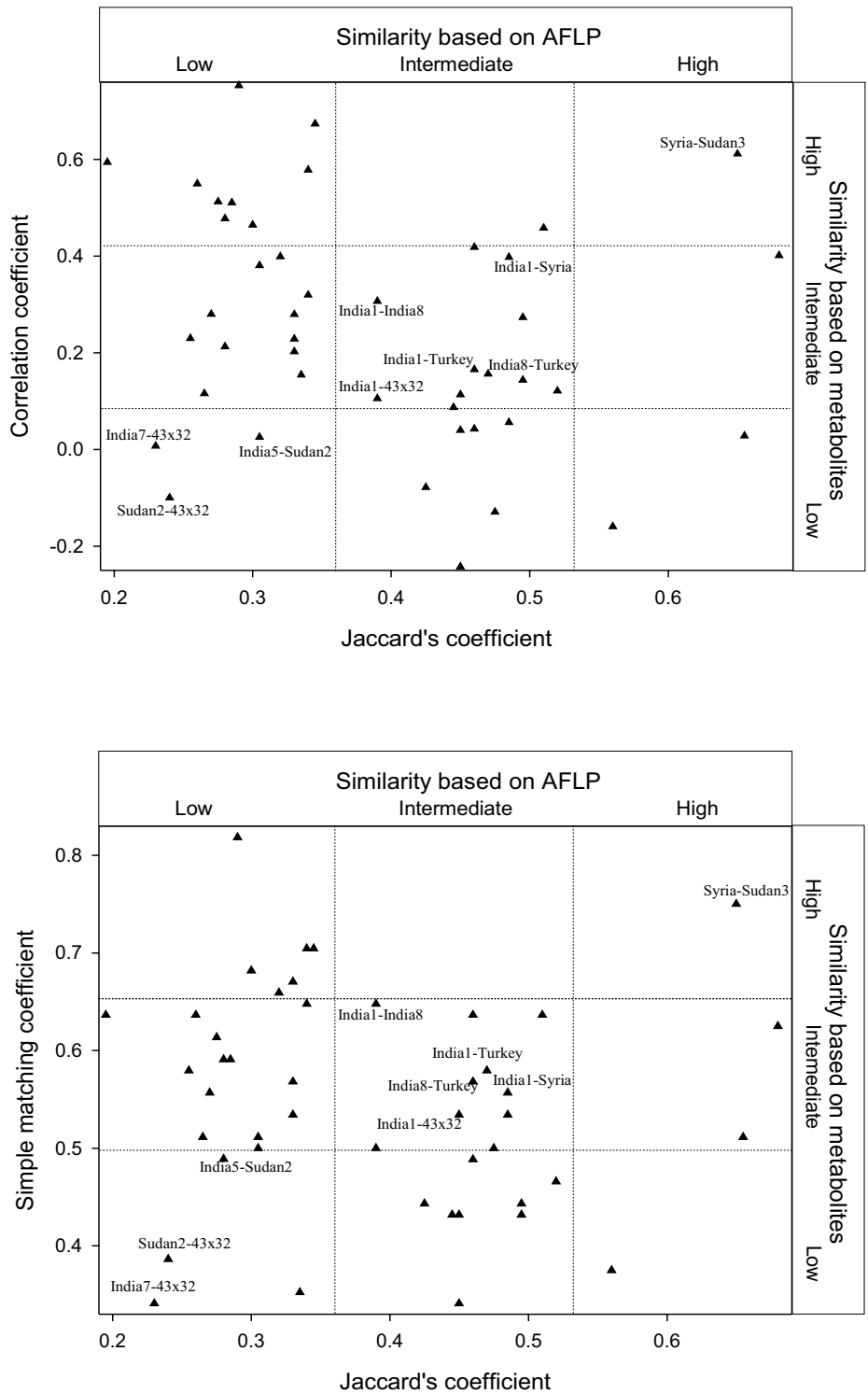

### Figure 1
**Scatter plots comparing ordination based on AFLP (Jaccard's coefficient) with ordination based on metabolic profiles**. Upper part: Metabolic profile comparisons based on quantitative variables (correlation coefficient). Lower part: Metabolic profile comparisons based on binary variables (simple matching coefficient). Accessions in pairwise comparisons which have a high, intermediate or low similarity for both approaches are labeled.

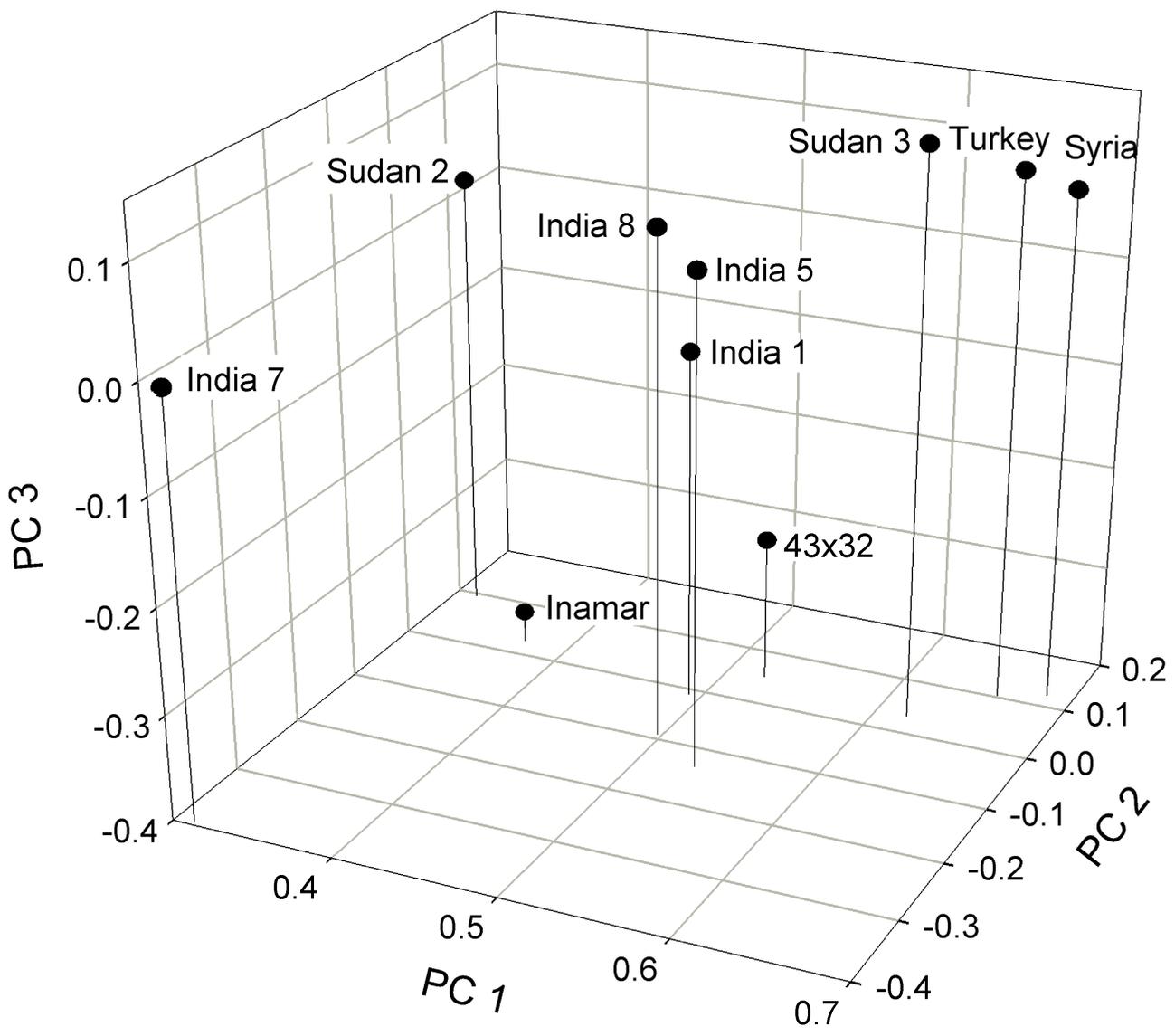

**Figure 2**
**Biplot of principal coordinate analysis based on Jaccard's coefficient for AFLP.**

ment of one accession in many pairwise comparisons would amplify this distortion. For example, two accessions in our set affect 17/45 pairwise comparisons. Thus a small number of biased data sets may alter the global pattern of biplots in a principal components or coordinates analysis. In this situation, scatter plots can identify which data sets are correlated, which are not, and which are not independent. Consistencies in scatter plots corroborate the representativeness of sampling in a particular pairwise comparison. For example, the pairs Sudan2-43 × 32, India7-43 × 32 and India5-Sudan2 were consistently the most dissimilar pairs in both the AFLP and the metabolic analysis. Similarly consistent were the comparison of pairs Syria-Sudan3 (highly similar for both approaches), and India1-Syria, India1-India8, India1-Turkey, India1-43 × 32 and I8-Turkey (intermediate similarity). Thus, the consistency of pairwise comparisons is independent of the similarity level.

Selection on the metabolome of a plant could distort the congruency in diversification between neutral DNA markers (AFLP) and metabolic profiles in a manner dependent on the intensity and duration of the selection pressure. Comparative analysis of intra- and interpopulation diversity at the genomic and metabolic levels will aid our understanding of the effect of selection on the evolution of metabolic capacity. Dedicated statistical tools that test the congruency in diversification of the metabolome and the genome are not available. However, tools for diversity estimation established in population genetics can be applied, offering at least qualitative insights.

Plants subjected to selection for metabolic traits should evolve faster on the metabolic level than neutral DNA markers should at the genomic level, as the rate of fixation of neutral mutations is controlled by only the mutation rate and population size. For example, Turkey-Syria accession pair appears to demonstrate the effect of selection on metabolic profiles of sesame (compare Fig. 2 and 3).

Common selection pressure exerted on different genotypes may result in different outcomes, i.e. convergent evolution or increased diversification. Increased diversification occurs when the biochemical basis of traits under selection differs among genotypes, e.g. when unrelated metabolic pathways enhance resistance to a common pathogen. The accession pair India7-Syria are very different at the genomic level but have similar metabolic phenotypes, and could have resulted from convergent evolution driven by common selection on genotypes with the same metabolic potential. Alternatively, neutral markers may have diversified over a long period of time, during which the metabolic phenotype was maintained by constant selection pressure.

The third situation encountered in our comparison of genomic and metabolic diversity in sesame was that the relative amount of diversification between the members of a pair was qualitatively similar at both the genomic and the metabolic levels. Thus pairs that were highly different at the genomic level also were highly different at the metabolic level. We suggest that varying selection and complex evolutionary histories might explain this kind of data. The analysis of the inheritance of metabolic patterns and of the association between metabolic and genomic markers might provide deeper insights. We have begun to generate segregating populations to address these questions.

The purpose of untargeted metabolic profiling in our work was to sample metabolic diversity without bias for the biological activity or practical relevance of the underlying compounds. One might want to know, however, whether metabolites of particular interest have been recovered in ethanol extracts used for the analysis. The most prominent metabolites of sesame are phenylpropanoids with one or more methylenedioxybenzole (piperonyl-) moieties such as sesamin and sesaminol. These lignans occur in free form and as di- and triglucosides and possess antioxidative properties. Certain sesame lignans lower blood and liver cholesterol levels, qualifying as health-promoting agents. In traditional analytical protocols, crushed sesame seeds are defatted with hexane prior extraction with ethanol or methanol. The defatting step is often used in lignan analysis in order to improve the recovery [42], but the lignans of sesame can be extracted from oil directly into methanol [43,44], indicating that defatting is not necessary. Indeed, an HPLC method for the analysis of sesame lignans based on extraction with 80% ethanol without defatting was described [45]. Similarly, extraction of sesame with methanol without defatting was used for sesamin determination [46]. In line with these results, we observed that the recovery of eight sesame lignans did not improve substantially by defatting seeds prior ethanol extraction (data not shown), which we used in the comparison of lignan content among sesame accessions [47]. As long as the life span of reverse phase columns is not a matter of concern, defatting seeds prior extraction can be omitted.

## Conclusion
Diversity patterns in sesame (*Sesamum indicum* L.) at the genomic level (neutral DNA markers) and at the metabolic level (nontargeted HPLC-MS profiles) differed, often showing a higher diversification rate at the metabolic level. For sesame breeders this means that the distances among accessions determined by genome fingerprinting need not reflect differences in metabolic capacity. Genetic analyses based on neutral markers is not an accurate predictor of the potential of parental lines for breeding programs aiming to improve traits controlled by metabolic phenotype such as resistance to pests or taste. The complementation of AFLP fingerprints by metabolic profiles for breeding and conservation purposes in sesame is recommended.

## Methods
### Plant material
Seeds were obtained from Centro Nacional de Investigaciones Agropecuarias (CENIAP) Germplasm Bank, Venezuela (Table 3). Plants were germinated and grown in the greenhouse with a photoperiod of 12 hours dark and 12 hours light at 30°C.

### AFLP analysis
DNA was extracted from leaves and AFLP analysis was performed based on the protocol by Voss et al. [2] with minor modifications as previously reported [25,26], using eight primer combinations (Table 1). AFLP reactions were performed twice for each accession, using restriction enzymes EcoRI and Tru1I (MBI Fermentas, Germany) and compatible primers (see Table 1 and Table 7 in [25]). Primers for pre-amplification were extended by one selective nucleotides (C for MseI and A for EcoRI). During selective amplification, fluorescent label (Cy5) was attached to the EcoRI primer. DNA fragments were separated on ALFexpress II DNA analyzer (Amersham Pharmacia Biotech, Uppsala, Sweden). Automatic band recognition and matching was done by using GelCompar II software (Applied Math, Belgium). A threshold value of 5% relative to the maximum value within each lane was applied and only fragments identified in both replicas (between 94 and 100% of all bands recorded) were used for band matching. The results of band matching were encoded as a binary matrix, which was used for all further analysis.

### Metabolic profiling
Seeds originating from five plants per accession were bulked and 1 g of tissue was frozen with liquid nitrogen, ground in a mortar with a pestle and extracted anaerobically with a mixture of 80% ethanol (gradient grade, Roth, Germany) and 20% water for 16 h with stirring (100 rpm). The liquid phase was filtered through 0.2 μm filters and kept at -20°C until HPLC analysis.

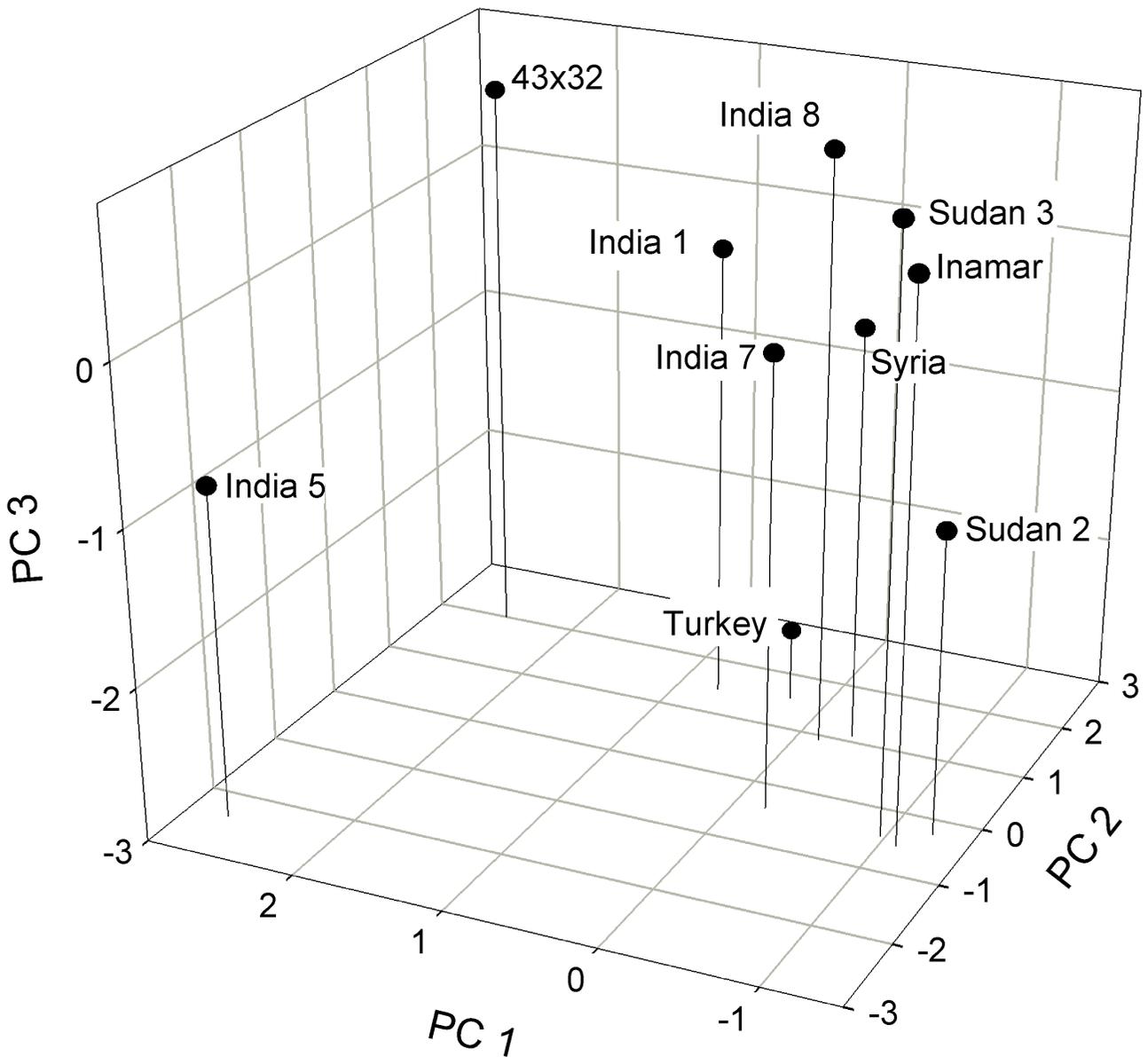

**Figure 3**
**Biplot of principal components analysis based on correlation coefficient for seed metabolic profiles**

**Table 3: Sesame accessions**

| CENIAP Germplasm Bank | | | |
|---|---|---|---|
| Accessions | Country of Origin | Working code | Diversity Centre |
| 93–2223 | India | India 1 | India |
| 89–007 | India | India 5 | India |
| 95–464 | India | India 7 | India |
| 92–2918 | India | India 8 | India |
| 92–2922 | Turkey | Turkey | Western Asia |
| 93–2022 | Syria | Syria | Western Asia |
| 92–310 | Sudan | Sudan 2 | Africa |
| 92–2872 | Sudan | Sudan 3 | Africa |

| Venezuelan accessions | | |
|---|---|---|
| Accessions | Country of Origin | Description |
| Experimental line 43 × 32 | Generated in Venezuela | Line selected from a second cycle of recurrent selection toward high yield under heavy whitefly infestation. The original population was obtained by cross, one to one, among 50 exotic accessions [48] |
| Commercial cultivar Inamar | Developed in Venezuela | Individual selection from the offspring from the same Acarigua's parents [49] |

For HPLC analysis, 10 μl aliquots of extracts were loaded onto a polar-modified RP-18 phase column (C18-Pyramid, Macherey-Nagel, Düren, Germany, 3 μm, 2 × 125 mm) and separated at 40°C with a gradient of 10% – 98% methanol at a flow rate of 0.2 ml min-1. The eluent was subjected to electrospray ionisation (ESI). Ions were analyzed in both positive and negative full scan mode between 50 and 1000 m/z with an ion trap.

*Data processing and analysis*
Raw data from the metabolic study were processed with the CODA algorithm (background reduction and spike elimination [33]). Extracted ion chromatograms with a mass quality index of at least 0.85 (according to technical manual of ACD/MS Manager v. 8.0, Advanced Chemistry Development, Toronto, Canada) were generated and compared. Based on these chromatograms, peak tables were generated. Ten peaks with the highest MCQ value for each accession were selected. For each peak, matching peaks in all accessions were identified, building a set of peaks for use in further analysis. Isotope peaks, recognized by the difference of one unit in the molecular weight and the same retention time, were combined to generate one value per metabolite per accession.

Peak areas were standardized twice, first within every accession by dividing the area by the total sum of areas of all peaks for each accession to compensate for loading differences, and second within every m/z value (across accessions) by dividing peak areas by the maximum area within the m/z value. The purpose of the second normalization was to weight major peaks in each extracted ion chromatogram equally for statistical evaluation, because the relationship between the amount of a substance that enters the ion source and the magnitude of the signal recorded by a mass detector varies among metabolites. Due to the lack of a suitable criterion, no data pretreatment was applied [34]. The resulting matrix was used to calculate correlation coefficients as a measure of similarity between pairs of accessions. To assess the effect of differences in signal intensities within extracted ion chromatograms, the matrix of doubly-normalized intensities was transformed into a binary matrix by replacing all nonzero values with 1. Using the binary matrix, a simple matching coefficient was calculated for each pair of accessions. The correlation between the correlation coefficient-based matrix and simple matching coefficient-based matrix was calculated by Mantel test (500 permutations). To visualize the relationship among accessions according to their metabolite content, principal component analysis was conducted with the correlation matrix. Principal coordinate analysis was used for the simple matching coefficient matrix. Calculations of similarity and dissimilarity coefficients, principal component and coordinate analysis were performed with NTSySpc 2.11T (Applied Biostatistics, Setauket (NY), USA).

A binary matrix from the AFLP data was obtained and a Jaccard's coefficient similarity matrix was calculated. The relationship among accessions was visualised as a principal coordinate analysis. Comparison of ordination obtained by AFLP and metabolite content was based on Pearson's correlation and a Mantel test between the matrices with 1000 permutations. The two approaches also were compared by scatter plots to visualize the correlation. The variability range in the scatter plots was split into three sections (high similarity, intermediate similarity and low similarity) on both the X axis and the Y axis. Pairwise comparisons for the same category in both approaches were identified i.e. pairs of accessions that were highly similar both in AFLP and metabolic data, pairs that possessed an intermediate similarity in both data sets, and pairs dissimilar both in genome and metabolome. The results of principal coordinate analysis performed on AFLP data and principal component analysis performed on metabolic data were compared visually.

## Authors' contributions

HL participated in the design of the study, conducted the genomic diversity study and statistical analysis, and drafted the manuscript. AR generated and analyzed metabolic profiles. PK conceived the study, participated in the experimental design and data analysis and wrote parts of the manuscript. All authors read and approved the final version of the manuscript.

## Acknowledgements

This work was supported by the Programme Alban, European Union Programme of High Level Scholarships for Latin America, International PhD program for Agricultural Sciences in Göttingen University (IPAG) and Universidad Centroccidental Lisandro Alvarado.

## References

1. Ovesná J, Poláková K, Leisová L: **DNA analyses and their applications in plant breeding.** *Czech J Genet Plant Breed* 2002, **38:**29-40.
2. Voss P, Hogers R, Bleeter M, Reijans M, Lee T van de, Hornes M, Frijters A, Pot J, Peleman J, Kuiper M, Zabeau M: **AFLP: a new technique for DNA fingerprinting.** *Nucleic Acids Res* 1995, **23:**4407-4414.
3. Li YC, Korol AB, Fahima T, Beiles A, Nevo E: **Microsatellites: Genomic distribution, putative functions and mutational mechanisms.** *Molecular Ecology* 2002, **11:**2453-65.
4. Dunn W, Ellis D: **Metabolomics: Current analytical platforms and methodologies.** *Trends in Analytical Chemistry* 2005, **24:**285-294.
5. Hall RD: **Plant metabolomics: From holistic hope, to hype, to hot topic.** *New Phytologist* 2006, **169:**453-468.
6. Fiehn O: **Metabolomics – the link between genotypes and phenotypes.** *Plant Molecular Biology* 2002, **48:**155-171.
7. Fiehn O, Weckwerth W: **Deciphering metabolic networks.** *Eur J Biochem* 2003, **270:**579-588.
8. Bedigian D, Harlan J: **Evidence for cultivation on sesame in the ancient world.** *Economic Botany* 1986, **40:**137-154.
9. Bedigian D: **Evolution of sesame revisited: domestication, diversity and prospects.** *Genet Res Crop Evol* 2003, **50:**779-787.
10. Yoshida H, Takagi S: **Effects of seed roasting temperature and time on the quality characteristics of sesame (*Sesamum indicum*) oil.** *J Sci Food Agric* 1997, **75:**19-26.
11. Shyu Y, Hwang L: **Antioxidative activity of the crude extract of lignan glycosides from unroasted Burma black sesame meal.** *Food Research International* 2002, **35:**357-365.
12. Dachtler M, Put F van de, Stijn F, Beindorff C, Fritsche J: **On-line LC-NMR-MS characterization of sesame oil extracts and assessment of their antioxidant activity.** *Eur J Lipid Sci Technol* 2003, **105:**488-496.
13. Shirato-Yasumoto S, Komeichi M, Okuyama Y, Horigane A: **A simplified HPLC quantification of sesamin and sesamolin in sesame seed.** *SABRAO Journal of Breeding and Genetics* 2003, **35:**27-34.
14. Kang MH, Naito M, Tsujihara N, Osawa T: **Sesamolin inhibits lipid peroxidation in rat liver and kidney.** *J Nutr* 1998, **128:**1018-1022.
15. Kang MH, Kawai Y, Naito M, Osawa T: **Dietary defatted sesame flour decreases susceptibility to oxidative stress in hypercholesterolemic rabbits.** *J Nutr* 1999, **129:**1885-1890.
16. Kang MH, Naito M, Sakai K, Uchida K, Osawa T: **Mode of action of sesame lignans in protecting low-density lipoprotein against oxidative damage in vitro.** *Life Sci* 2000, **66:**161-171.
17. Miyake Y, Fukumoto S, Okada M, Sakaida K, Nakamura Y, Osawa T: **Antioxidative catechol lignans converted from sesamin and sesaminol triglucoside by culturing with *Aspergillus*.** *J Agric Food Chem* 2005, **53:**22-27.
18. Devine G, Denholm I: **An unconventional use of piperonyl butoxide for managing the cotton whitefly, *Bemisia tabaci* (Hemiptera:Aleyrodidae).** *Bulletin of Entomological Research* 1988, **88:**601-610.
19. Devine G, Ishaaya I, Horowitz A, Denholm I: **The response of pyriproxyfen-resistant and susceptible *Bemisia tabaci* Genn. (Homoptera:Aleyrodidae) to pyriproxyfen and fenoxycarb alone and in combination with piperonyl butoxide.** *Pestic Sci* 1999, **55:**405-411.
20. Brooker N, Long J, Stephan S: **Field assessment of plant derivative compound for managing fungal soybean diseases.** *Biochemical Society Transactions* 2000, **28:**917-920.
21. Victor S, Crisóstomo F, Bueno F, Pagnocca F, Fernandes J, Correa A, Bueno O, Hebling J, Bacci M Jr, Vieira P, daSilva F: **Toxicity of synthetic piperonyl compounds to leaf-cutting ants and their symbiotic fungus.** *Pest Manag Sci* 2001, **57:**603-608.
22. Harmatha J, Nawrot J: **Insect feeding deterrent activity of lignans and related phenylpropanoids with a methylenedioxyphenyl (pyperonyl) structure moiety.** *Entomologia Experimentalis et Applicata* 2002, **104:**51-60.
23. Nascimento I, Murata A, Bortoli S, Lopes L: **Insecticidad activity of chemical constituents from *Aristolochia pubescens* against *Anticarsia gemmatalis* larvae.** *Pest Manag Sci* 2004, **60:**413-416.
24. Peñalvo J, Heinonen S, Aura A, Adlerreutz H: **Dietary sesamin in converted to enterolactone in humans.** *J Nutr* 2005, **135:**1056-1062.
25. Laurentin H, Karlovsky P: **Genetic relationship and diversity in a sesame (*Sesamum indicum* L.) germplasm collection using amplified fragment length polymorphism (AFLP).** *BMC Genetics* 2006, **7:**10.
26. Laurentin H, Karlovsky P: **AFLP fingerprinting of sesame (*Sesamum indicum* L.) cultivars: identification, genetic relationship and comparison of AFLP informativeness parameters.** *Genetic Resources and Crop Evolution* 2007, **54:**1437-1446.
27. Wolde-meskel E, Terefework Z, Lindstrom K, Frostegard A: **Metabolic and genomic diversity of rhizobia isolated from field standing native and exotic woody legumes in southern Ethiopia.** *Syst Appl Microbiol* 2004, **27:**603-611.
28. Seymour F, Cresswell J, Jack P, Lappin-Scott H, Haag H, Talbot N: **The influence of genotypic variation on metabolite diversity in populations of two endophytic fungal species.** *Fungal Genetics and Biology* 2004, **41:**721-734.
29. Keurentjes JJ, Fu J, de Vos CH, Lommen A, Hall RD, Bino RJ, Plas LH van der, Jansen RC, Vreugdenhil D, Koornneef M: **The genetics of plant metabolism.** *Natural Genetics* 2006, **38:**842-849.
30. De Vos RC, Moco S, Lommen A, Keurentjes JJ, Bino RJ, Hall RD: **Untargeted large-scale plant metabolomics using liquid chromatography coupled to mass spectrometry.** *Nature Protocols* 2007, **2:**778-791.
31. Dixon RA, Gang DR, Charlton AJ, Fiehn O, Kuiper HA, Reynolds TL, Tjeerdema RS, Jeffery EH, German JB, Ridley WP, Seiber JN: **Applications of metabolomics in agriculture.** *J Agric Food Chem* 2006, **54:**8984-8994.
32. Rao N: **Plant genetic resources: advancing conservation and use through biotechnology.** *African Journal of Biotechnology* 2004, **3:**136-145.
33. Windig W, Phalp JM, Payne AW: **Noise and Background Reduction Method for Component Detection in Liquid Chromatography/Mass Spectrometry.** *Anal Chem* 1996, **68:**3602-3606.
34. Berg RA, Hoefsloot HCJ van den, Westerhuis JA, Smilde AK, Werf MJ, van der: **Centering, scaling, and transformations: improving the biological information content of metabolomics data.** *BMC Genomics* 2006, **7:**142.
35. Joy N, Abraham Z, Soniya EV: **A preliminary assessment of genetic relationships among agronomically important cultivars of black pepper.** *BMC Genetics* 2007, **8:**42.

36. Bhattacharya E, Ranade SA: **Molecular distinction amongst varieties of mulberry using RAPD and DAMD profiles.** *BMC Plant Biology* 2001, **1**:3.
37. Awasthi AK, Nagaraja GM, Naik GV, Kanginakudru S, Thangavelu K, Nagaraju J: **Genetic diversity and relationships in mulberry (genus *Morus*) as revealed by RAPD and ISSR marker assays.** *BMC Genetics* 2004, **5**:1.
38. Dangi RS, Lagu MD, Choudhary LB, Ranjekar PK, Gupta VS: **Assessment of genetic diversity in *Trigonella foenum-graecum* and *Trigonella caerulea* using ISSR and RAPD markers.** *BMC Plant Biol* 2004, **4**:13.
39. Bhattacharya E, Dandin SB, Ranade SA: **Single primer amplification reaction methods reveal exotic and indigenous mulberry varieties are similarly diverse.** *J Biosci* 2005, **30**:669-77.
40. Tam SM, Mhiri C, Vogelaar A, Kerkveld M, Pearce SR, Grandbastien MA: **Comparative analyses of genetic diversities within tomato and pepper collections detected by retrotransposon-based SSAP, AFLP and SSR.** *Theor Appl Genet* 2005, **110**:819-31.
41. **Secondary Metabolites in Soil Ecology.** In *Soil Biology Volume 2*. Edited by: Karlovsky P, Varma A. New York: Springer; 2007:1-19.
42. Willför SM, Smeds AI, Holmbom BR: **Chromatographic analysis of lignans.** *J Chromatography* 2006, **1112**:64-77.
43. Dachtler M, Put FHM van de, Stijn Fv, Beindorff CM, Fritsche J: **On-line LC-NMR-MS characterization of sesame oil extracts and assessment of their antioxidant activity.** *Eur J Lipid Sci Technol* 2003, **105**:488-96.
44. Lim JS, Adachi Y, Takahashi Y, Ide T: **Comparative analysis of sesame lignans (sesamin and sesamolin) in affecting hepatic fatty acid metabolism in rats.** *Br J Nutr* 2007, **97**:85-95.
45. Yasumoto SS, Komeichi M, Okuyama Y, Horigane A: **A simplified HPLC quantification of sesamin and sesamolin in sesame seed.** *SABRAO J Breed Gen* 2003, **35**:27-34.
46. Williamson KS, Morris JB, Pye QN, Kamat CD, Hensley K: **A survey of sesamin and composition of tocopherol variability from seeds of eleven diverse sesame (*Sesamum indicum* L.) genotypes using HPLC-PAD-ECD.** *Phytochem Anal* 2007 in press. DOI: 10.1002/pca.1050
47. Hettwer U, Laurentin H, Karlovsky P: **Determination of antioxidative furofuran lignans in sesame seeds by HPLC-MS.** In *Proceedings of the Second International Symposium on Recent Advances in Food Analysis, Last Minute Posters: Prague* Edited by: Hajslova J, Nielen MWF. Abingdon: Taylor and Francis; 2006:15. 2–4 November 2005
48. Laurentin H, Layrisse A, Quijada P: **Evaluación de dos ciclos de selección recurrente para altos rendimientos de semilla en una población de ajonjolí.** *Agronomía Tropical (Maracay)* 2000, **50**:521-535.
49. Mazzani B: **Inamar: nueva variedad de ajonjolí producida en el Instituto Nacional de Agricultura.** *Agronomía Tropical (Maracay)* 1953, **3**:211-213.